\pdfoutput=1
\documentclass[a4paper,11pt,notitlepage]{article}
\usepackage[textwidth=16.31cm, footskip=20mm, bottom=50mm]{geometry}
\usepackage{amsmath, amsfonts, amssymb}
\usepackage{amsthm}
\usepackage{authblk}
\usepackage{graphicx,color}
\usepackage{hyperref}
\usepackage{gensymb}
\usepackage{xspace}
\usepackage{upgreek}
\usepackage[super,comma,sort&compress,numbers]{natbib}

\newcommand{\supp}{Supplement\xspace}
\newcommand{\nm}{\,\mathrm{nm}}
\newcommand{\nA}{\,\mathrm{nA}}
\newcommand{\pA}{\,\mathrm{pA}}
\newcommand{\T}{\,\mathrm{T}}
\newcommand{\V}{\,\mathrm{V}}
\newcommand{\mV}{\,\mathrm{mV}}
\newcommand{\eV}{\,\mathrm{eV}}
\newcommand{\meV}{\,\mathrm{meV}}
\newcommand{\aF}{\,\mathrm{aF}}
\newcommand{\EF}{$E_\mathrm{F}$\xspace}
\newcommand{\Vtip}{$V_{\mathrm{tip}}$\xspace}
\newcommand{\Phigr}{$\Phi_\mathrm{gr}^\mathrm{el}$\xspace}

\newcommand{\Phizero}{$\Phi_0^\mathrm{el}$\xspace}

\begin{document}


\title{Electrostatically confined monolayer graphene quantum dots with orbital and valley splittings}	

\author[1]{Nils M.~Freitag}
\author[2]{Larisa A.~Chizhova}
\author[1]{Peter Nemes-Incze}
\author[3]{Colin R.~Woods}
\author[3]{Roman V.~Gorbachev}
\author[3]{Yang Cao}
\author[3]{Andre K.~Geim}
\author[3]{Kostya S.~Novoselov}
\author[2]{Joachim Burgd\"orfer}
\author[2]{Florian Libisch}	
\author[1, *]{Markus Morgenstern}
\affil[1]{II. Institute of Physics B and JARA-FIT, RWTH Aachen University, Otto-Blumenthal-Stra{\ss}e, 52074 Aachen, Germany}
\affil[2]{Institute for Theoretical Physics, TU Wien, Wiedner Hauptstra{\ss}e 8-10, 1040 Vienna,
  Austria, EU}
\affil[3]{School of Physics \& Astronomy, University of Manchester, Manchester, United Kingdom}
\affil[*]{corresponding author: MMorgens@physik.rwth-aachen.de}

\maketitle

\begin{abstract}
The electrostatic confinement of massless charge carriers 
	is hampered by Klein tunneling. 
Circumventing this problem in graphene mainly 
	relies on carving out nanostructures or applying 
	electric displacement fields to open a band gap in 
	bilayer graphene.
So far, these approaches suffer from edge disorder or 
	insufficiently controlled localization of electrons. 
Here we realize an alternative strategy in monolayer
	graphene, by combining a homogeneous magnetic field 
	and electrostatic confinement.
Using the tip of a scanning tunneling microscope, we 
	induce a confining potential in the Landau gaps of 
	bulk graphene without the need for physical edges.
Gating the localized states towards the Fermi energy 
	leads to regular charging sequences with more than 40 
	Coulomb peaks exhibiting typical addition energies 
	of $7$-$20\meV$. 
Orbital splittings of $4$-$10\meV$ and a valley 
	splitting of about 3$\meV$ for the first orbital state
	can be deduced.
These experimental observations are quantitatively 
	reproduced by tight binding calculations,
	which include the interactions of the graphene
	with the aligned hexagonal boron nitride substrate.
The demonstrated confinement approach appears suitable to create
	quantum dots with well-defined wave function properties 
	beyond the reach of traditional techniques. 
\end{abstract}

The charge carriers in graphene at low energies, 
	described as massless Dirac quasiparticles 
	\cite{CastroNeto2009}, are expected to 
	feature long 
	spin coherence times 
	\cite{Trauzettel2007,Fuchs2012,Droth2013,Fuchs2013}.
Exploiting this property requires precise 
	manipulation of individual Dirac electrons.
Quantum dots (QDs) present an essential building 
	block, yet providing tailored confinement in 
	graphene has remained challenging.
So far, e-beam lithography \cite{Bischoff2015} and
	various other techniques \cite{Giesbers2008, 
	Morgenstern2016,Magda2014,Liu2013, Qi2015,
	Vicarelli2015} have been used to design nanometer 
	sized devices.
However, their performance lacks behind, for 
	example, GaAs QDs \cite{Tarucha1996, Kouwenhoven2001}, as 
	disordered sample edges of patterned graphene 
	result in uncontrolled charge localization and 
	scattering \cite{Libisch2012, Bischoff2015, 
	Bischoff2016, Terres2016}. 
So far, no clear evidence for fourfold 
	degenerate charging sequences has been reported 
	in transport measurements of tunable QDs.
Moreover, failing to controllably lift 
	graphene's valley degeneracy renders spin qubits 
	unfeasible \cite{Trauzettel2007,Rycerz2007,Recher2009}.
\par
Bilayer graphene could, in principle, improve the situation, 
	since an electric displacement
	field opens a band gap at regular AB stacking \cite{Zhang2009}.
Indeed, electrostatically confined QDs in bilayer graphene
	exhibit Coulomb blockade
	\cite{Allen2012,Goossens2012,Muller2014}, yet controlling 
	the spin or valley degree of freedom of an 
	individual state has also not been demonstrated. 
Moreover, confinement is still prone to parasitic 
	conduction channels due to residual disorder in the 
	band gap or conducting channels along domain walls 
	of AB- and BA-stacked areas \cite{Ju2015}.
Another approach exploits whispering gallery modes 
	in electrostatically confined QDs 
	\cite{Zhao2015,Lee2016,Gutierrez2016}, 
	but here the control of the wave functions by 
	gates is difficult and dwell times are extremely 
	short ($<100\,\mathrm{fs}$).
On an even more intricate route, the tip of a scanning 
	tunneling microscope (STM) is used to locally 
	stretch a suspended monolayer graphene sheet 
	\cite{Klimov2012}.
The onset of charge quantization due to induced strain 
	showcases confinement by pseudomagnetic fields.
Adding a real magnetic field $B$ leads to charging 
	sequences with regular orbital but no valley 
	splittings \cite{Klimov2012}.
Creating multiple QDs in this fashion would require 
	independent strain control for every QD on the 
	suspended graphene.
Thus such an approach is barely scalable.	
\par
Landau quantization helps to overcome Klein tunneling 
	by opening band gaps \cite{Allen2012, Goossens2012, Muller2014}.
An elegant method to exploit this by combining a 
	magnetic field and an electrostatic potential 
	has been proposed theoretically 
	\cite{Chen2007,Giavaras2009,Giavaras2012}. 
Indeed, indications of such confinement have
	been found in metal contact induced 
	pnp junctions \cite{Moriyama2014}, graphene on 
	$\text{SiO}_2$ \cite{Jung2011, LuicianMayer2014} 
	and a suspended graphene nano-ribbon 
	\cite{Tovari2016}.
However, in these experiments the confinement 
	potential was not tunable but was generated by
	electrostatic disorder.
\par
\begin{figure*}[htbp]
    \includegraphics{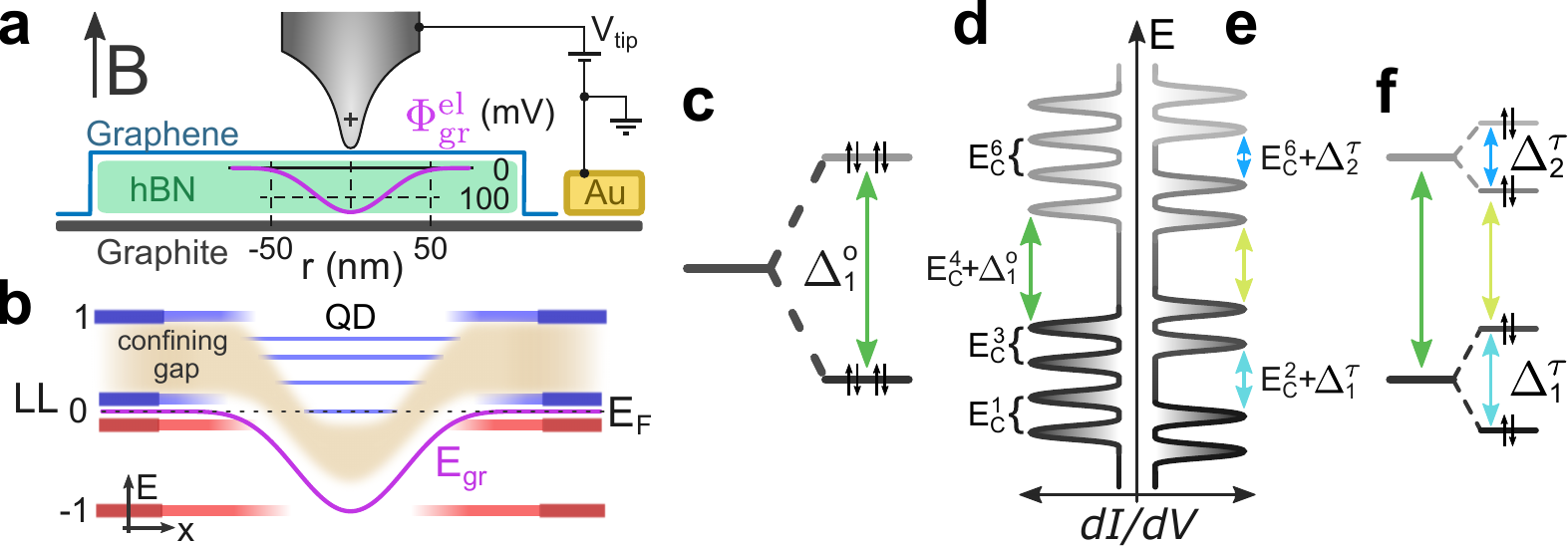}
    \caption{
			\textbf{(a)} Sketch of the experiment. Graphene 
				covers a $30\nm$ thick hexagonal boron nitride
				flake on graphite.
				The magenta line represents the tip-induced confinement 
				potential of graphene \Phigr for electrons, 
				calculated as the numerical solution of Poisson's 
				equation (\supp).
      \textbf{(b)} Energy diagram in real space: 
				Fermi energy \EF, black dashed line; local
				band bending $E_\mathrm{gr}$, magenta line;
				states belonging to electron (hole) LLs, blue 
				(red); bulk LLs, 1, 0, -1.
				States embedded in the LL$_0$-LL$_{+1}$ gap (thin blue lines)
				are electrostatically confined.
			\textbf{(c)} Energy level diagram for the first two orbital 
				states of a graphene QD exhibiting an orbital splitting 
				$\Delta^\mathrm{o}_1$. Both orbitals are fourfold 
				degenerate, as indicated by black arrows representing physical spin.
			\textbf{(d,e)} Charging peak sequence in the differential 
				conductance $dI/dV$ corresponding to the level diagrams
				in \textbf{c} and \textbf{f}, respectively. 
				Charging peaks are separated by the addition energy 
				$E_\mathrm{add}^i = E_\mathrm{C}^i + \Delta_i$, where
				$E_\mathrm{C}^i \approx E_\mathrm{C}$ is the charging energy and 
				$\Delta_i$ is comprised of $\Delta^\mathrm{o}_j$ and/or
				the valley splittings $\Delta^\mathrm{\tau}_k$. 
				In \textbf{d} quadruplet ordering showcases a dominant 
				$\Delta^\mathrm{o}_1$, while  
				$\Delta^\mathrm{\tau}_k$ become sizable in \textbf{e},
				further separating quadruplets into doublets.
			\textbf{(f)} Same as \textbf{c}, but including additional 
				$\Delta^\mathrm{\tau}_k$. The spin splitting 
				$\Delta^\sigma$ is neglected, as 
				$\Delta^\sigma < \Delta^\mathrm{\tau}_k, \Delta^\mathrm{o}_j,
				E_\mathrm{C}$ in experiment.
			}
		\label{fig:Zero}
\end{figure*}

%
Here, we demonstrate controlled confinement by a combination of
	magnetic and electrostatic fields.
We use the tip-induced electrostatic potential of an
  STM \cite{Dombrowski1999,Morgenstern2000} 
	in a $B$ field perpendicular to the graphene plane 
	(Fig.~\ref{fig:Zero}a). 
Scanning tunneling spectroscopy (STS) reveals sequences of 
	charging peaks by means of Coulomb staircases which 
	appear when these confined states cross the Fermi energy 
	\EF.
The peaks systematically group in quadruplets
  for electrons and holes corresponding to the fourfold (valley and
  spin) degeneracy in graphene (Fig.~\ref{fig:Zero}c,d).
Moreover, some quadruplets separate into doublets due to an
	additional valley splitting induced by the 
	hexagonal boron nitride (BN) substrate.
STS as a function of $B$ reveals that the first confined states
	emerge from Landau levels (LLs) with indices $\pm1$.
A third-nearest neighbor tight binding (TB) 
	calculation \cite{Libisch2010,Chizova2014} reproduces 
	the onset of charging events as function of tip 
	voltage \Vtip and $B$, and the magnitude of 
	orbital and valley splittings.
\par
%
We now sketch the principle of our experiment.
A homogeneous, perpendicular $B$ field condenses the electronic
states of graphene into LLs at energies
\begin{equation}\label{eq:Landau}
	E_N = \mathrm{sgn}\left(N\right)\sqrt{2\hbar e \nu_\text{F}^2\left|BN\right|},
\end{equation}
where $\nu_\mathrm{F}$ is the Fermi velocity and $N\in\mathbb{Z}$ is the LL index \cite{CastroNeto2009}. 
Consequently, energy gaps between the LLs emerge in the electronic 
	spectrum.  
The smooth electrostatic potential \Phigr 
	(magenta line in Fig.~\ref{fig:Zero}a) induced 
	by the STM tip locally shifts the eigenenergies 
	$\varepsilon_i (\Phi_\mathrm{gr}^\mathrm{el})$ of
	charge carriers relative to the bulk LL energy
	(eq~\ref{eq:Landau}). 
Shifting $\varepsilon_i$ into the Landau gaps 
	creates confined states (Fig.~\ref{fig:Zero}b) 
	\cite{Giavaras2009}. 
The shape of \Phigr determines the single-particle orbitals
	and energy levels, as in the case of artificial atoms 
	\cite{Kouwenhoven2001}.
Orbital splittings $\Delta^\mathrm{o}_j$ separate 
	the energy levels (Fig.~\ref{fig:Zero}c), 
	which we deduce experimentally to be 
	$\Delta^\mathrm{o}_j = 4-10\meV$ (see below)
	and, thus, $\Delta^\mathrm{o}_j$ is small compared 
	to the first LL gap $E_1-E_0 \approx 100\meV$ at $7\T$. 
While pristine graphene exhibits a fourfold 
	degeneracy, varying stacking orders of graphene 
	on top of BN induce an additional valley splitting 
	$\Delta^\mathrm{\tau}_k$, which turns out to be 
	smaller than $\Delta^\mathrm{o}_j$ in our experiment.
The finite $B$ field creates a small Zeeman 
	splitting estimated as
	$\Delta^\sigma = g\mu_\mathrm{B}B\approx 
	800\,\mathrm{\upmu eV}$ at $7\T$ 
	($g$-factor of 2, $\mu_\mathrm{B}$: Bohr's magneton). 
Accordingly, the orbital splittings separate quadruplets 
	of near-degenerate QD states, which exhibit a subtle 
	spin-valley substructure (Fig.~\ref{fig:Zero}f).
\par
We use the STM tip not only as source of the electrostatic 
	potential and thus as gate for the QD states but also to
	sequence the energy level spectrum of the QD as the 
	states cross \EF, that is, as the charge on the QD 
	changes by $\pm e$.
This leads to a step in the tunneling current
	$I(V_{\mathrm{tip}})$ and a corresponding charging peak
	in the differential conductance $dI/dV_{\mathrm{tip}}$.
In addition to the single particle energy spacings, 
	every additional electron on the dot needs to overcome 
	the electrostatic repulsion to the electrons already 
	inside the QD \cite{Averin1986}, given by the charging 
	energy $E_\mathrm{C}^i$. 
Thus, we probe the total energetic separation of charge 
	states $i$ and $i+1$, given by the addition energy 
	$E_\mathrm{add}^i = E_{\mathrm C}^i + \Delta_i$,
	where $\Delta_i$ consists of
	$\Delta^\mathrm{o}_j$, $\Delta^\mathrm{\tau}_k$ and/or
	$\Delta^\mathrm{\sigma}$.
As we experimentally find $E_{\mathrm{C}}^i \approx 
	E_{\mathrm{C}} \approx 10\meV 
	\gtrsim \Delta^\mathrm{o}_j$ (nearly independent of the 
	charge state $i$, see below),
	the quadruplet near-degeneracy of the QD states translates
	to quadruplet ordering of the charging peaks 
	(Fig.~\ref{fig:Zero}d).
Whenever either $\Delta^\mathrm{\tau}_k$ or $\Delta^\mathrm{\sigma}$ 
	significantly exceeds the other and temperature, quadruplets separate 
	into doublets (Fig.~\ref{fig:Zero}e).
\par
%
%
We prepare our sample (see  Fig.~\ref{fig:Zero}a and \supp)
	by dry-transferring \cite{Mayorov2011,Kretinin2014}
	a graphene flake onto BN
	\cite{Gorbachev2014,Hunt2013,Woods2014}.
During this step we align both crystal lattices with 
	a precision better than one degree (\supp).
Then we place this graphene/BN stack on a large 
	graphite flake to avoid insulating areas and simplify 
	navigating the STM tip.
Any disorder potential present in the sample 
	will limit the confinement as long as it is larger 
	than the Landau level gaps, thus larger gaps (e.g., the 
	\mbox{LL$_0$ - LL$_{\pm1}$} gap) result in 
	improved confinement. 
Moreover, the induced band bending will only be well-defined 
	if the disorder potential is smaller than 
	the maximum of \Phigr.
By using the dry-transfer technique
	\cite{Mayorov2011,Kretinin2014} and a graphite/BN 
	substrate we reduce disorder in the graphene 
	significantly \cite{Decker2011,Xue2011,Deshpande2012}.
\par
%
\begin{figure*}[htp]
    \includegraphics{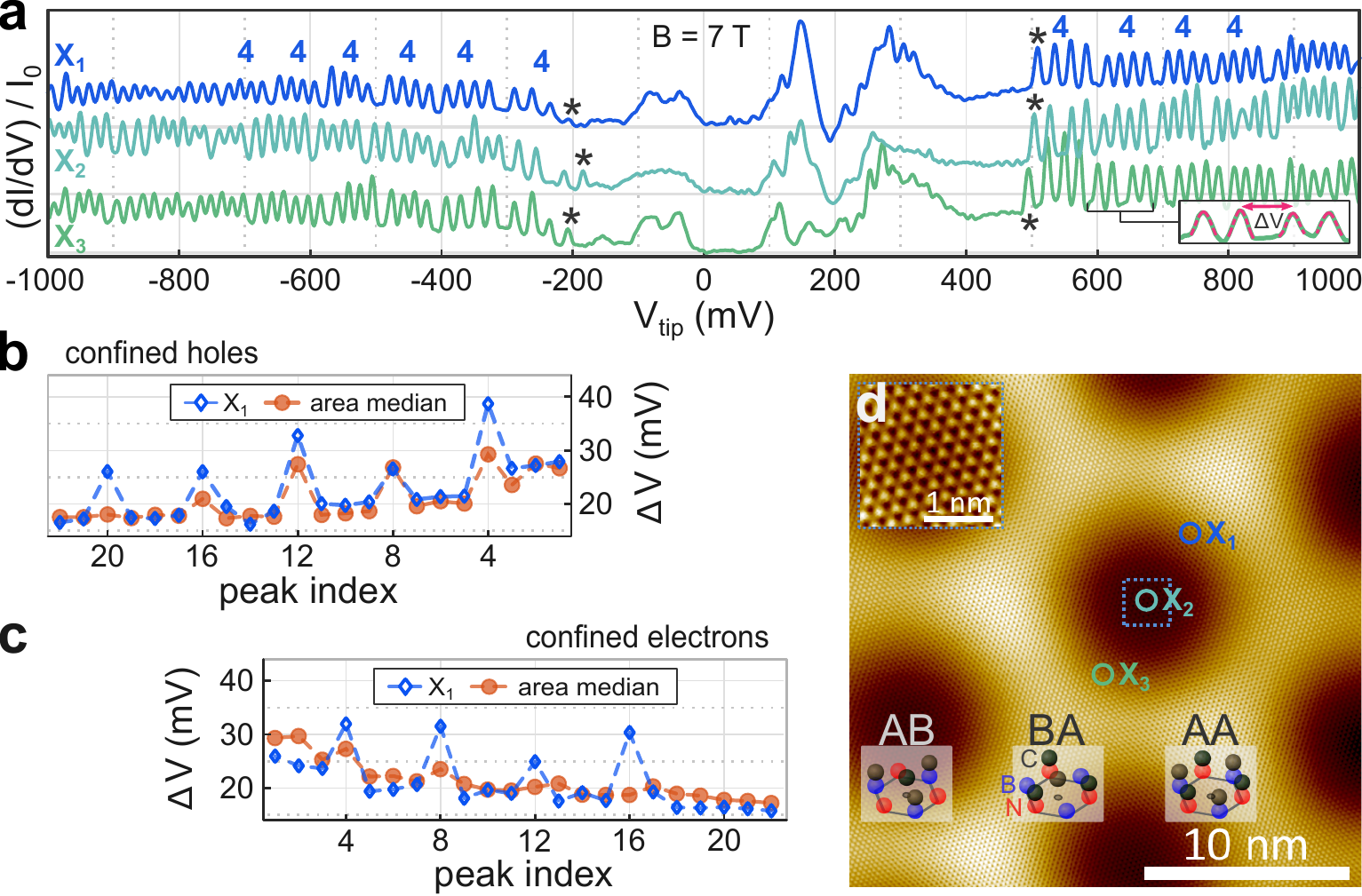}
    \caption{
			\textbf{(a)} Representative differential conductance 
				spectra $dI/dV(V_\mathrm{tip})$, normalized by the first 
				value $I_0$ of the respective $I(V_\mathrm{tip})$ curve 
				(\supp).
				Recording positions are: $\mathrm{X}_1$, 
				between AA and AB; $\mathrm{X}_2$, on AB; $\mathrm{X}_3$, 
				between AB and BA (compare \textbf{d}).
				Spectra on other regions (e.g., AA, BA) look similar. 
				$V_\mathrm{stab}=1\V$, $I_\mathrm{stab}=700\pA$, 
				$V_\mathrm{mod} = 4.2\mV_\mathrm{rms}$ and $B=7\T$. 
				Quadruplets of peaks are marked by $`` 4 "$ and the 
				first charging peak on either \Vtip side by an asterisk.
				Curves are offset for clarity, while horizontal gray lines mark 
				$dI/dV=0\,\mathrm{S}$. 
				Inset shows a zoom with Gaussian fits (dashed lines)
				used to extract distances between adjacent peaks $\Delta V$
				as marked.
			\textbf{(b,c)} $\Delta V$ as function of 
				consecutive peak index, for spectrum $\mathrm{X}_1$ 
				(blue, error bars smaller than symbol size) and the 
				median values for $80\times80$ spectra recorded on 
				$60\times60\nm^2$ (orange).
			\textbf{(d)} Atomically resolved STM image (raw data)
				of the aligned graphene on hexagonal boron nitride (BN). 
				$V_\mathrm{tip} = 400\mV$, $I = 1\nA$. 
				Differently stacked areas AB, BA and AA marked and 
				sketched by ball models.
				Inset on the upper left shows a zoom into the AB stacked area, 
				marked by the blue square, exhibiting an obvious sublattice
				symmetry breaking due to the underlying BN.
				Positions equivalent to those where spectra in \textbf{a}
				were recorded are marked by circles labeled
				$\mathrm{X}_1$,$\mathrm{X}_2$,$\mathrm{X}_3$.
			}
		\label{fig:One_topo_model_QD_evidence}
\end{figure*}
Probing the sample in our custom-build 
	UHV-STM system \cite{Mashoff2009} at $T=8\,$K,
	we observe the superstructure with $a= 13.8\nm$ 
	periodicity, which develops due to the small 
	lattice mismatch of $1.8\%$ 
	between graphene and BN \cite{Xue2011}.
An atomically resolved STM image of this 
	superstructure is presented in 
	Figure~\ref{fig:One_topo_model_QD_evidence}d.
Prior to measuring $dI/dV$ spectra, the tip-sample 
	distance is adjusted at the stabilization voltage 
	$V_\mathrm{stab}$ and current $I_\mathrm{stab}$ 
	and then the feedback loop is turned off (\supp).
Figure \ref{fig:One_topo_model_QD_evidence}a 
	shows exemplary $dI/dV$ spectra, acquired at 
	$B=7\T$ and adjusted to same vertical scale by dividing
	$dI/dV$ by the first value $I_0$ of the respective $I(V)$
	curve (\supp).
We observe pronounced, regularly spaced 
	peaks for $V_\mathrm{tip} < -170\mV$ and 
	$V_\mathrm{tip} > 500\mV$. 
A closer look at the sequences reveals the expected
	grouping in quadruplets,
	which can still be distinguished up to the 
	$20^{\mathrm{th}}$ peak.
This grouping becomes even more evident by directly comparing
	the voltage difference between adjacent peaks $\Delta V$ 
	in Figure~\ref{fig:One_topo_model_QD_evidence}b,c:
	$\Delta V$ between quadruplets is up to twice as large
	as $\Delta V$ within the quadruplets indicating 
	$\Delta^\mathrm{o}_j \lesssim E_\mathrm{C}^i$ while 
	$\Delta^\mathrm{\tau}_k$ and $\Delta^\mathrm{\sigma}$
	are significantly smaller.
To further elucidate grouping patterns, we 
	measure 6400 $dI/dV$ spectra at equidistant 
	positions within a $60\nm\times60\nm$ area,
	thus probing all areas of the superstructure.
The median $\Delta V$ values (orange circles in 
	Fig.~\ref{fig:One_topo_model_QD_evidence}b,c)
	portray the robust ordering into quadruplets 
	on the hole side, implying $\Delta^\mathrm{o}_j$
	generally dominates over $\Delta^\tau_k$ and 
	$\Delta^\mathrm{\sigma}$.
On the electron side of the spectra the sequences are 
	disturbed by a few additional 
	charging peaks of defect states in the BN substrate 
	\cite{Wong2015} which are identified 
	by their characteristic spatial development (\supp).
This limits the comparability of the electron and hole 
	sector and hides possible smaller electron-hole 
	asymmetries in the data.
The $dI/dV$ features in between the charging peaks most 
	likely capture contributions from multiple orbital states 
	of each LL, which are lifted in degeneracy by the 
	tip-induced potential, but cannot be identified 
	unambiguously (\supp, Sec.~5).
\par
%
\begin{figure}[htbp]
		\centering
			\includegraphics[]{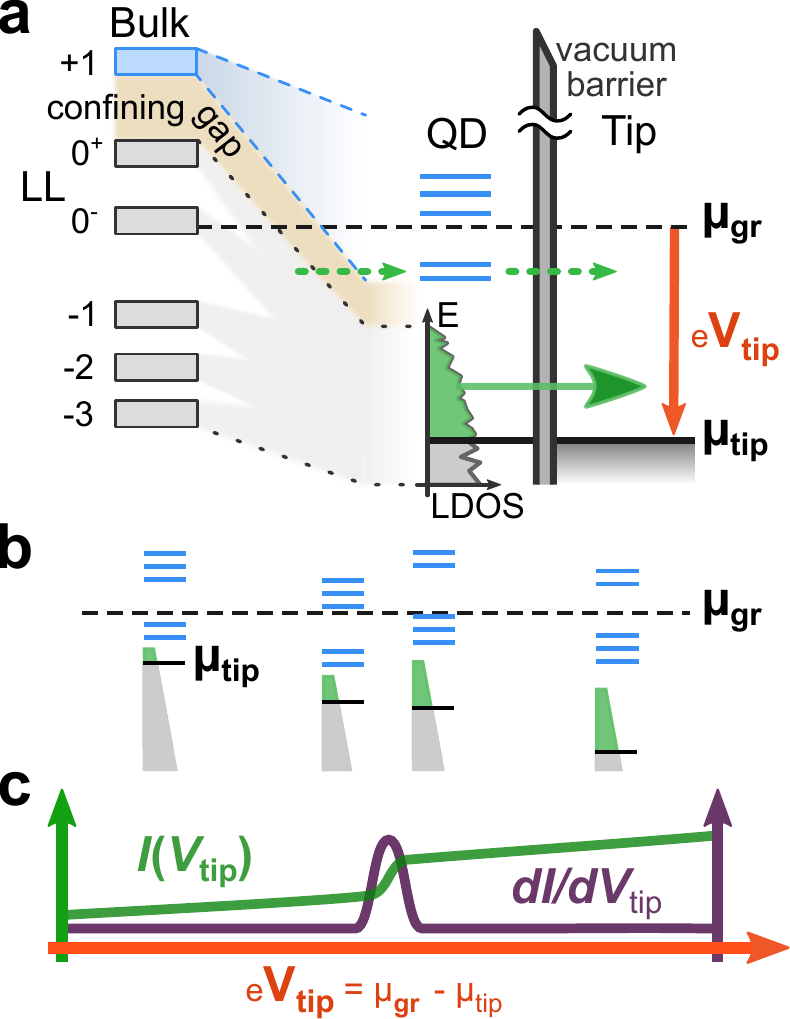}
    \caption{ Sketch of the Coulomb staircase.
		\textbf{(a)} 
			The chemical potentials of graphene $\mu_\mathrm{gr}$ 
			(black dashed line) and tip $\mu_\mathrm{tip}$
			(black solid line) define the bias window 
			$eV_\mathrm{tip}$, within which 
			graphene states tunnel into empty tip states. 
			There are two current paths available: 
			(i) a weak one (green dashed arrow) via quantum dot 
			states (blue lines),
			(ii) a dominant one (solid green arrow) via states 
			strongly coupled to the graphene bulk (marked LDOS).
			Left: bulk graphene LLs away from the tip-induced
			band bending.
		\textbf{(b)} Schematic diagram of change in QD energies (blue 
			lines) and quasi-continuous LDOS underneath the tip
			(green and gray triangle) for increasing $V_\mathrm{tip}$ 
			from left to right. 
			Between the second and third frame, the QD changes 
			its charge state shifting
			the energy of the QD states and the entire LDOS upwards. 
		\textbf{(c)} Tunneling current $I$ displaying the staircase 
			(green line) and 
			differential conductance $dI/dV$ (purple line) for 
			increasing \Vtip (aligned with \textbf{b}).
			}
	\label{fig:Two_CP_model}
\end{figure}
To understand the origin of the charging peaks, 
	we provide a detailed microscopic picture 
	of the tip-induced gating of localized states.
We will only discuss the case of positive
	\Vtip, that is, electron confinement, since 
	the arguments for negative \Vtip are analogous.
Increasing \Vtip (orange arrow in 
	Fig.~\ref{fig:Two_CP_model}) shifts the 
	states underneath the tip energetically down.
States originating from LLs with positive index are 
	embedded in the \mbox{LL$_0$-LL$_{+1}$ gap}
	which provides electrostatic confinement
	(Fig.~\ref{fig:Two_CP_model}a, 
	see also Fig.~\ref{fig:Zero}b). 
Within the bias window $eV_\mathrm{tip} = 
	\mu_\mathrm{gr} - \mu_\text{tip}$, electrons 
	tunnel from the sample into unoccupied states 
	of the tip.
One current path (dashed green arrow 
	Fig.~\ref{fig:Two_CP_model}a) passes through 
	states of the QD (blue lines).
The other stronger current path (solid green arrow 
	Fig.~\ref{fig:Two_CP_model}a) originates from the 
	quasi-continuous LDOS at lower energies 
	where energetically overlapping LL states  
	strongly couple to the graphene bulk.
Though increasing \Vtip gates QD states down 
	(Fig.~\ref{fig:Two_CP_model}b), 
	the	Coulomb gap around \EF always separates the 
	highest occupied from the lowest unoccupied state,
	prohibiting continuous charging of confined states. 
It is only when the next unoccupied level crosses 
	$\mu_\mathrm{gr}$ that the QD is charged by
	an additional electron.
The electrostatic repulsion due to its charge 
	abruptly increases the Hartree energy
	of all states, thereby shifting additional
	graphene states from below $\mu_\mathrm{tip}$
	into the bias window (Fig.~\ref{fig:Two_CP_model}b, 
	central transition).
Consequently, the tunneling current $I$ increases
	which translates to a charging peak in 
	$dI/dV_\mathrm{tip}$ (Fig.~\ref{fig:Two_CP_model}c).
This mechanism is called Coulomb staircase 
	\cite{Averin1986} and has been observed previously, 
	for instance, for charging of clusters within an 
	STM experiment \cite{Hanna1991}.
In essence, charging peaks in $dI/dV$ signal the 
	coincidence of a charge level of the QD with 
	$\mu_\mathrm{gr}$ \cite{Wildoer1996} and thus provide a 
	clear signature of the addition energy spectrum of the QD.
\par
Since the measurement captures the QD level spacings 
	as charging peak distances $\Delta V$, they need to 
	be converted to $E_\mathrm{add}$ via the tip lever 
	arm $\alpha_\text{tip}$.
The latter relates a change of \Vtip to its induced
	shift of the QD state energies.
The lever arm is determined by the ratio of the capacitance
	between tip and dot $C_\mathrm{tip}$, and the total 
	capacitance of the dot $C_\Sigma$, thus 
	$\alpha_\text{tip} = C_\mathrm{tip}/C_\mathrm{\Sigma}$.
$C_\mathrm{\Sigma}$ includes $C_\mathrm{tip}$,
	the capacitance between dot and back-gate, and dot and 
	surrounding graphene.
We use a Poisson solver to estimate 
	\mbox{$C_\mathrm{\Sigma} = 16.5\pm3.2\aF$} and 
	\mbox{$C_\mathrm{tip} = 8\pm1.5\aF$} for our QD
	(\supp).	
Hence, we find $E_\mathrm{C} = e^2/C_\mathrm{\Sigma}
	\approx 10\pm2\meV$ and  
	\mbox{$\alpha_\text{tip} = 0.51\pm0.03$} (close to values 
	reported for a similar system by \citet{Jung2011}).	
Consequently charging peaks dominantly separated by 
	$E_\mathrm{C}$, that is, 
	\mbox{$E_\mathrm{add}^i \approx E_\mathrm{C}^i$}
	because \mbox{$\Delta_i \ll E_\mathrm{C}^i$}, should exhibit
	\mbox{$\Delta V = E_\mathrm{C}/
	(e\cdot\alpha_\mathrm{tip})\approx 20\mV$}, in close
	agreement with the values found within quadruplets
	at higher occupation numbers
	(Fig.~\ref{fig:One_topo_model_QD_evidence}b,c).
As expected, we also find significantly larger 
	$E_\mathrm{add}^i$ for every fourth charging peak.
In case of clear quadruplet ordering, the orbital 
	splittings for our QD are deduced from 
	$\Delta^\mathrm{o}_j = 
	E_\mathrm{add}^{4j} - E_\mathrm{C}^{4j}
	\approx E_\mathrm{add}^{4j} - E_\mathrm{add}^{4j+1}$
	and we find typical values of $4-10\meV$ for the
	first few orbitals ($\alpha_\mathrm{tip} = 0.51$,
	Fig.~\ref{fig:One_topo_model_QD_evidence}b,c).
For this estimate we neglect the additional Zeeman splitting
	or an even smaller valley splitting. 
\par
We next provide a theoretical framework to elucidate
	the details of the QD level spectrum. 
The eigenstates of bulk graphene LLs (eq~\ref{eq:Landau}) 
	feature different wave function amplitudes
	on sublattices\cite{CastroNeto2009} A and B,
\begin{equation} 
\Psi_N^\mathrm{K} = \begin{pmatrix} \Psi^\mathrm{A}_{|N|-1} \\ \Psi^\mathrm{B}_{|N|} \end{pmatrix} 
	\hspace{10pt} \mathrm{and}\hspace{12pt} 
\Psi_N^\mathrm{K'} = \begin{pmatrix} \Psi^\mathrm{A}_{|N|} \\ \Psi^\mathrm{B}_{|N|-1} \end{pmatrix},
\label{eq:WF-LLs}
\end{equation}
	where K and K' denote the two inequivalent K-points of 
	the Brillouin zone associated with the two valleys.
For $N \ne 0$ the LL index differs by one for the two 
	sublattices, while for $N = 0$ the part of the wave 
	function with subscript $\left| N \right|-1$ 
	vanishes, resulting in polarized 
	sublattices for each valley.  
The wave functions of bulk graphene (eq~\ref{eq:WF-LLs}) 
	are modified by the tip-induced potential.  
Assuming a radially symmetric confinement potential, 
	the eigenstates are described by radial and angular
	momentum quantum numbers $\left(n_\mathrm{r}, m \right)$,
	with $n_\mathrm{r}\in \mathbb{N}_0 $ and $m\in \mathbb{Z}$.
Adiabatically mapping a given LL with index $N$ on to
	possible combinations of $n_\mathrm{r}$ and $m$ yields 
	\cite{Schnez2008,Yoshioka2007}
\begin{equation} \label{eq:N_restriction}
	|N| = n_\mathrm{r} + 1/2\cdot\left( m + \left| m \right|\right), 
\end{equation}
	with $0\leq n_\mathrm{r}\leq |N| $ and 
	$m \leq |N|$.
\par
\begin{figure*}[htp]
    \centering
    \includegraphics[width=\textwidth]{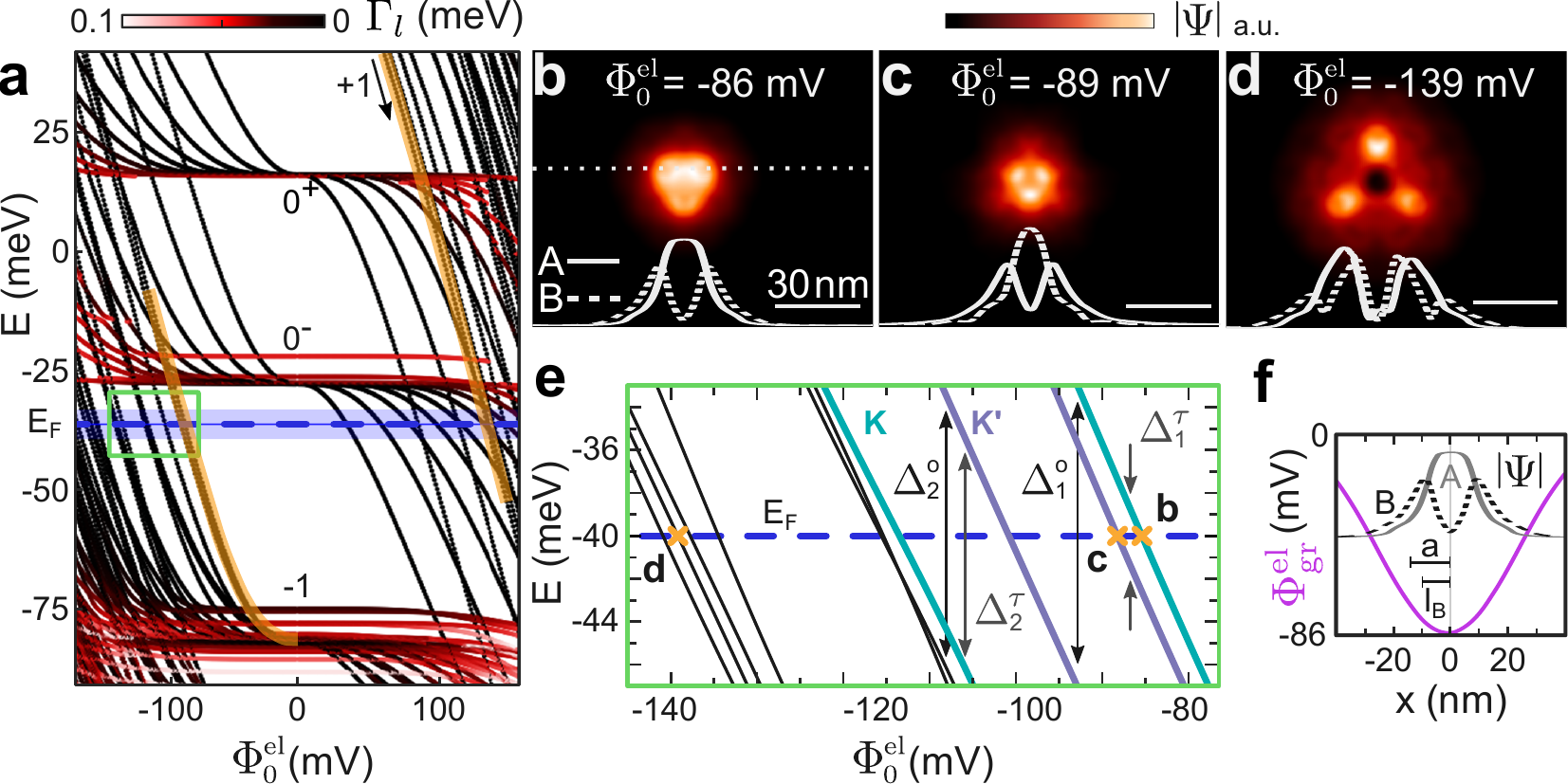}
    \caption{ 
		\textbf{(a)} Tight binding eigenenergies of a 
			$120\times100\nm^2$ graphene sample with open boundaries 
			as function of tip-induced potential amplitude \Phizero at 
			$B=7\T$ with the tip-induced potential centered on an AB
			area (BA and AA yield very similar behavior, not shown). 
			Line color encodes coupling to the boundary 
			(imaginary part $\Gamma_l$ of eigenenergies): 
			black (red) indicates strong (weak) 
			localization underneath the tip.
			States from LL$_{\pm1}$ and the split LL$_0$ are 
			labeled by $\pm1$ and 0, respectively.
			The LL$_0$ splitting reduces the confining gap to
			$E_{0^-}-E_{-1} \approx 50\meV$.
			First states crossing \EF from LL$_{\pm1}$ are 
			highlighted in orange.
			Uncertainty in $E_\mathrm{F}$ indicated as blue 
			horizontal bar (\supp).
			The green rectangle marks the zoom shown in \textbf{e}.
		\textbf{(b-d)} Color plot of the wave function amplitude $|\Psi|
			= \sqrt{|\psi_\mathrm{A}|^2 + |\psi_\mathrm{B}|^2}$ of states marked 
			by orange crosses in \textbf{e}. \Phizero at 
			the crossing point 
			$\varepsilon_l(\Phi_0^\mathrm{el}) = E_\mathrm{F}$ 
			is marked. 
			Solid (dashed) white lines are line cuts 
			along the dotted white line in \textbf{b} for 
			contributions from sublattice A (B), as marked. 
			All scale bars identical.
		\textbf{(e)} Zoom into area marked by a green box in 
			\textbf{a}. 
			Colored lines identify valley K (cyan) and K' (purple). 
			Orange crosses mark crossing of \EF (blue dashed line)
			of selected states, 
			which are displayed in \textbf{b}-\textbf{d}.
			First two orbital $\Delta^\mathrm{o}_j$ and valley 
			$\Delta^\mathrm{\tau}_k$ splittings marked by arrows.
		\textbf{(f)} Comparison of length scales: 
			tip-induced potential, magenta; calculated wave 
			function amplitude $|\Psi|$ of first 
			state crossing \EF (same as \textbf{b}) for sublattice 
			A (gray line) and B (dashed line); superstructure lattice 
			constant $a = 13.8\nm$; magnetic length 
			$l_\text{B}(7\T) = 9.7\nm$.
		}
\label{fig:Three_dot_states_V0_dependence}
\end{figure*}

We calculate eigenstates of a $120\nm\times100\nm$ commensurate
	graphene flake on BN using third-nearest neighbor TB 
	\cite{Libisch2010}, where the substrate interaction enters via 
	a periodic superstructure potential and local strain 
	effects \cite{Chizova2014}, parametrized from DFT 
	calculations \cite{Sachs2011,Martinez2014}.
We approximate the amplitude $\Phi_0^\mathrm{el}$ 
	and shape of \Phigr by a classic electrostatic 
	solution of Poisson's equation (Fig.~\ref{fig:Zero}a, 
	\supp) with the tip radius $r_\mathrm{tip}$ as fit parameter.
Comparing calculated charging energies to experiment 
	yields a plausible value of $r_\mathrm{tip} \approx 120\nm$
	implying a FWHM of the QD confinement potential 
	of $55\nm$ at $7\T$.
We independently determine the initially free parameter \EF 
	from the position of LL$_0$ in STS as \mbox{\EF$=-40\pm5\meV$} 
	(\supp).
Accordingly, the graphene is p-doped.
We note that varying \EF within the stated uncertainty range
	(see blue horizontal bar in
	Figure~\ref{fig:Three_dot_states_V0_dependence}a)
	leads to no qualitative changes in the 
	predictions of our model.
We use open boundary conditions to simulate the coupling of the
	flake to the surrounding graphene. 
Consequently, eigenstates will feature complex eigenvalues 
	$E_l = \varepsilon_l + i \Gamma_l/2$, where the real part 
	$\varepsilon_l$ represents the resonant energies and 
	the imaginary part $\Gamma_l$ the coupling to the 
	delocalized bulk states \cite{Bischoff2014}. 
Thus we can readily distinguish states that are
	spread out over the flake (large $\Gamma_l$) from 
	those localized near the tip (small $\Gamma_l$).
We color code $\Gamma_l$ in 
	Figure~\ref{fig:Three_dot_states_V0_dependence}a for a
	calculation with the tip-induced potential centered on 
	an AB stacked area.
\par
%
At $B=7\T$ and vanishing band bending 
	($\Phi_0^\mathrm{el}=0$), we find only delocalized 
	states whose eigenenergies cluster around the bulk LL 
	energies (eq~\ref{eq:Landau},
	Fig.~\ref{fig:Three_dot_states_V0_dependence}a).
As we increase \Phizero, states begin to localize at the 
	tip and shift in energy, with smaller $\Gamma_l$ 
	(darker curves) pointing to stronger localization (see 
	Fig.~\ref{fig:Three_dot_states_V0_dependence}a).
Comparing hole states originating from LL$_{-1}$ for 
	negative and positive \Phizero, we find, as expected, 
	stronger localization in case of negative \Phizero.  
The potential is always attractive to one kind of charge 
	carriers which will localize underneath the tip.
The other kind is repelled by the 
	induced potential (see also Ref.~\citenum{Giavaras2012})
	which results in stronger coupling to the bulk.
In order to classify our TB wave functions in terms of 
	the quantum numbers $N$, $n_\mathrm{r}$ and $m$, we 
	consider sublattice A and B separately.
Tracing the states back to their LL of origin reveals $N$, 
	constraining possible $n_\mathrm{r}\le |N|$. 
The value of $n_\mathrm{r}$ is then determined by counting 
	radial minima in the line cuts of the 
	wave function amplitude for each sublattice
	(Fig.~\ref{fig:Three_dot_states_V0_dependence}b-d). 
The distance of the first radial maximum from the center of
	the wave function is finally sufficient to assign the 
	possible $m$ quantum numbers of the LL (eq~\ref{eq:N_restriction}).
Additionally, the ($n_\mathrm{r},m$) combinations need to
	be consistent with $N$ differing by one on the two 
	sublattices (eq~\ref{eq:WF-LLs}).
For instance, the line cuts in 
	Figure~\ref{fig:Three_dot_states_V0_dependence}b
	portray $(0,0)$ and $(0,1)$ on sublattice A and B, 
	respectively. 
As expected, small angular momentum states are the first 
	ones to localize with increasing \Phizero, in line 
	with calculations by \citet{Giavaras2009}.
Notice that the applied $B$ naturally lifts the orbital 
	degeneracy in QDs \cite{Fock1928}. 	
Delocalized states remain at bulk LL energies
	(red horizontal lines in 
	Fig.~\ref{fig:Three_dot_states_V0_dependence}a).
\par
We distinguish two regimes in the sequence of spin degenerate
	states crossing \EF for negative \Phizero.
The first regime (Fig.~\ref{fig:Three_dot_states_V0_dependence}e)
	exhibits $ \Delta^\mathrm{\tau}_k \lesssim \Delta^\mathrm{o}_j
	\lesssim E_\mathrm{C}^i$,
	while the second at higher \Phizero is characterized by 
	densely spaced states, thus $\Delta^\mathrm{o}_j \approx 
	\Delta^\mathrm{\tau}_k \ll E_\mathrm{C}^i$.
The sequence within the first regime corresponds to about five 
	orbital pairs from valley K and K', in line with about five 
	quadruplets in our experimental spectra  (see labels ``4" in 
	Fig.~\ref{fig:One_topo_model_QD_evidence}a and 
	$\Delta V$ sequences in 
	Fig.~\ref{fig:One_topo_model_QD_evidence}b,c).
The quite uniform spacing of peaks for larger \Vtip 
	(Fig.~\ref{fig:One_topo_model_QD_evidence}a) 
	agrees with the second regime. 
In order to extract $\Delta^\mathrm{o}_j$ and 
	$\Delta^\mathrm{\tau}_k$ within the first regime,
	we carefully assign the valley index to the  
	states.
Using the previously determined $n_\mathrm{r}$ and $m$
	in eq~\ref{eq:N_restriction},
	the first state crossing \EF
	(Fig.~\ref{fig:Three_dot_states_V0_dependence}b) 
	features LL index $N_\mathrm{A}= 0 + 1/2(0+|0|) = 0$ 
	on sublattice A and $N_\mathrm{B}= 1 + 1/2(0+|0|) = 1$ 
	on sublattice B, as predicted by eq~\ref{eq:WF-LLs} 
	for a LL$_{|1|}$ state in valley K.
The role of the sublattices interchanges for
	the second state crossing \EF 
	(Fig.~\ref{fig:Three_dot_states_V0_dependence}c),
	placing it in valley K'.
Consequently, states with $N_\mathrm{A}=N_\mathrm{B}-1$ and 
	$N_\mathrm{B}=N_\mathrm{A}-1$ are assigned to valleys K 
	and K', respectively.
The calculation therefore predicts a valley splitting
	of about $\Delta_1^\mathrm{\tau}= 3\meV$ on the AB and BA 
	areas (see 
	Fig.~\ref{fig:Three_dot_states_V0_dependence}b,c,e).
$\Delta_2^\mathrm{\tau}$ is comparatively large 
	(about $12\meV$) and the respective orbital splitting 
	$\Delta_2^\mathrm{o}$ is only larger by $1-2\meV$
	(see Fig.~\ref{fig:Three_dot_states_V0_dependence}e).
Consequently additional electrons may occupy the next orbital state of
	one valley prior to the same orbital state of the 
	other valley at higher occupation numbers.
Hence we limit further comparison to experiment to 
	$\Delta_1^\mathrm{\tau}$.
In our TB model, the strength of the valley splitting 
	is dominated by the sublattice symmetry breaking term 
	due to the BN substrate \cite{Chizova2014}.
The calculations also show that the radial extent of 
	the wave functions grows for the first couple of 
	states crossing \EF, as expected for increasing $m$
	(compare Fig.~\ref{fig:Three_dot_states_V0_dependence}b,c
	to Fig.~\ref{fig:Three_dot_states_V0_dependence}d), 
	explaining the decrease of 
	$E_{\mathrm{C}}^i$ towards higher peak indices at fixed $B$
	(see Fig.~\ref{fig:One_topo_model_QD_evidence}b,c).
\par
%
\begin{figure*}

    \centering
    \includegraphics{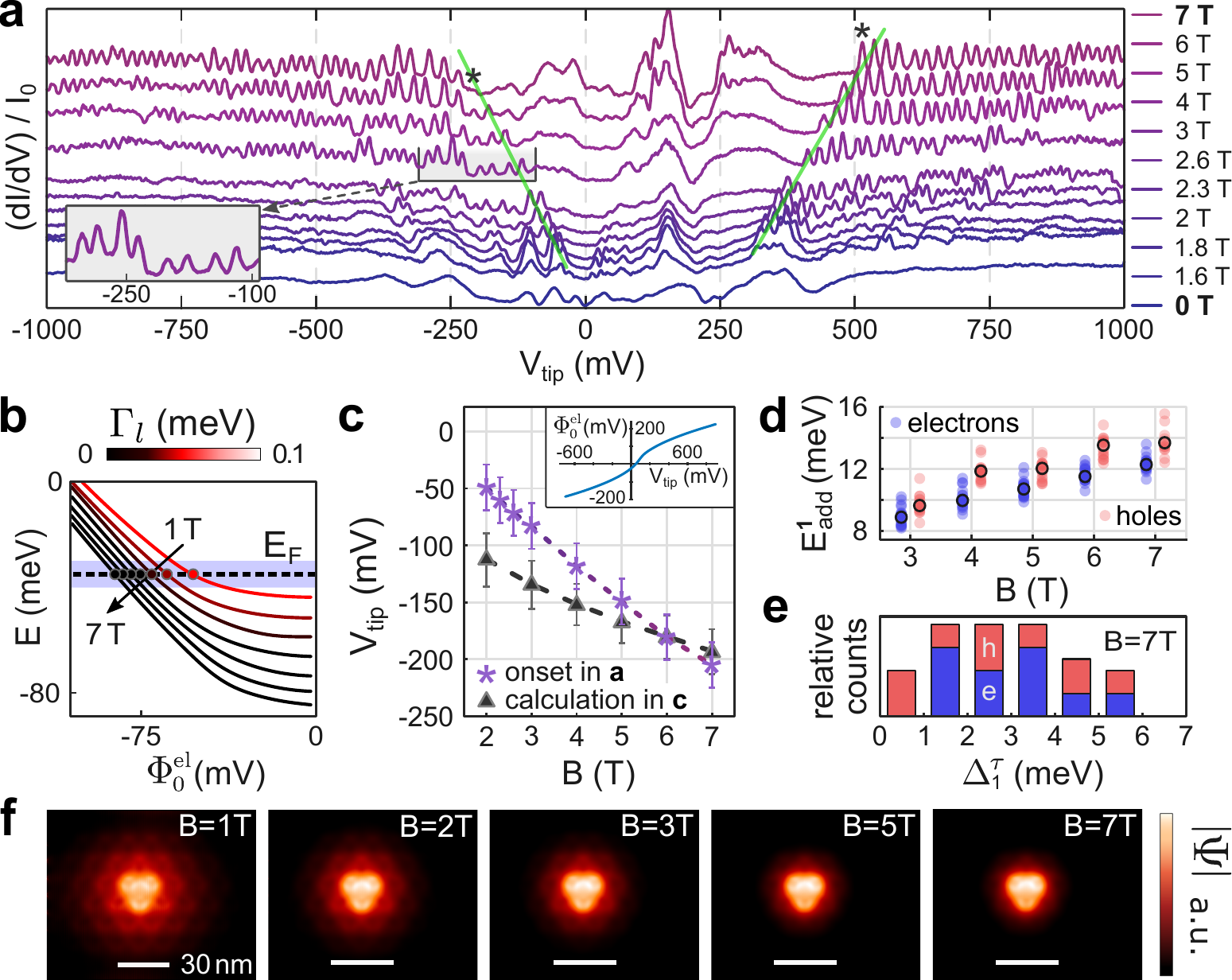}

    \caption{ 
		\textbf{(a)} $dI/dV$ spectroscopy in the vicinity of 
			an AA stacked area at varying $B$, marked on the right.
			Four spatially adjacent spectra are averaged 
			and the ones for 
			$B>0\T$ are offset by a value proportional to $\sqrt{B}$. 
			\mbox{$V_\mathrm{stab} = 1000\mV$}, 
			\mbox{$I_\mathrm{stab} = 700\pA$}, 
			\mbox{$V_\mathrm{mod} = 4.2\mV_\mathrm{rms}$}.
			Green lines are guides to the eye, marking the onset voltage
			of charging peaks $V^*$. At $7\T$ an asterisk marks the first 
			charging peak on either side.
			Inset shows zoom onto marked peaks. 
		\textbf{(b)} Energy of first confined hole state as function 
			of induced potential amplitude \Phizero for different 
			$B$ as marked.
			At larger $B$, states cross \EF at larger \Phizero,
			shifting $V^*$ to larger negative \Vtip. 
			Color codes imaginary part of the eigenenergy as in 
			Fig.~\ref{fig:Three_dot_states_V0_dependence}a.  
		\textbf{(c)} Comparison between measured and calculated $V^*$. 
			Inset shows the required $\Phi_0^\mathrm{el}(V_\mathrm{tip})$
			for conversion, taken from a Poisson-solver (\supp)
			and including a reasonable work function difference of 
			$\Delta\Phi = 50\meV$ between graphene and tip. 
			Error bars for measured $V^*$ reflect typical variation 
			of $V^*$ on AA areas across a few superstructure unit cells. 
			Error bars for calculation arise from the 
			uncertainty in \EF.
		\textbf{(d)} Plot of the $B$ dependance of $E_\mathrm{add}^1 
			\approx E_\mathrm{C}^1$
		  of 20 spectra (semitransparent dots) in the vicinity of an AA area.
			Data points are recorded at integer valued $B$ fields (in Tesla),
			but displayed
			slightly shifted to the left (electrons, blue) and to the right
			(holes, red) for clarity. Median values are encircled in black. 
		\textbf{(e)} Histogram of
			$\Delta^\tau_1 \approx E^2_\mathrm{add}-E^3_\mathrm{add}$ 
			(experimental error below $0.2\meV$) at $B=7\T$ 
			for the same AA area used in \textbf{d}.  
			Electron (blue bars) and hole 
			(red bars) contributions are colored. 
		\textbf{(f)} Calculated $|\Psi|$
      of the first confined hole state (see \textbf{b}) crossing 
			\EF at different 
			$B$ as marked. The state originates from LL$_{-1}$,
			i.e., \Vtip$<0\V$ when crossing. All scale bars identical.
			}
\label{fig:Four_mag_field_dependency}
\end{figure*}
%
Theory and experiment can be directly compared for
	the $B$ dependence of the onset voltage 
	of charging peaks $V^*$. 
Experimentally, $V^*$ shifts towards higher
	$\left| V_\mathrm{tip} \right|$ for increasing $B$
	(Fig.~\ref{fig:Four_mag_field_dependency}a),
	thus gating the first state to \EF requires stronger 
	band bending for higher $B$.
Since the curves for $B>0\T$ are offset proportional to 
	$\sqrt{B}$, the straight line connecting the first 
	charging peaks reveals that the energy distance of 
	the first state to \EF scales with $\sqrt{B}$.
This corresponds to the increase in bulk LL energies 
	for $N\neq 0$ (eq~\ref{eq:Landau}), strongly suggesting
	those LLs as source of the confined states.
This analysis is confirmed by our TB calculations, as 
	the first crossing points of LL$_{\pm1}$ states 
	with the Fermi level $\Phi^*_0$ also shift towards
	higher $\left| \Phi_\mathrm{0}^\mathrm{el} \right|$
	with increasing B
	(Fig.~\ref{fig:Four_mag_field_dependency}b).
While the evolution of states with \Phigr in 
	Figure~\ref{fig:Three_dot_states_V0_dependence}a
	is (approximately) symmetric with respect to 
	\mbox{\Phigr$\rightarrow -$\Phigr},
	the previously discussed p-doping induces an
	asymmetry in $\Phi^*_0$ for electrons and holes
	(see the lines highlighted in orange in
	Fig.~\ref{fig:Three_dot_states_V0_dependence}a)
	and thus accounts for the observed asymmetry in $V^*$. 
In Figure~\ref{fig:Four_mag_field_dependency}c 
	we compare $V^*$ and $\Phi^*_0$ by using the
	$\Phi_0^\mathrm{el}(V_\mathrm{tip})$ dependence 
	from the Poisson solver 
	(see inset Fig.~\ref{fig:Four_mag_field_dependency}c,
	\supp). 
Care must be taken to correctly account for the work 
	function difference between the tip and the sample:
	the tip's work function ($4.5-4.8\eV$ \cite{Dombrowski1999, 
	Chen2016}) exceeds that of graphene ($4.5\eV$), placing 
	electric field neutrality in the positive \Vtip sector.
Moreover, it definitely has to lie in between the two 
	charging peak regimes because the QD vanishes without 
	band bending. 
Using a plausible work function difference of $+50\meV$ in
	Figure~\ref{fig:Four_mag_field_dependency}c leads to  
	satisfactory agreement between the theoretical predictions
	for the first state crossings and the experimental $V^*$.
\par
Our TB simulations predict a strong reduction of 
	$\Gamma_l$ with increasing magnetic field, 
	corresponding to the suppression of the
	radial tail of the wave function in 
	Figure~\ref{fig:Four_mag_field_dependency}f
	and indicating the onset of localization 
	between 1 and $3\T$
	(Fig.~\ref{fig:Four_mag_field_dependency}b).
The first appearance of charging peaks in the experiment 
	at around $2\T$ 
	(Fig.~\ref{fig:Four_mag_field_dependency}a) 
	fits nicely.
This finding is further corroborated by comparing the 
	diameter of the LL state 
	$d_n = 2\sqrt{2n+1}\cdot l_\mathrm{B}$, being 
	\mbox{$d_1 = 89\nm$}~$(63\nm)$ for LL$_1$ at $1\T~(2\T)$,
	with the FWHM of the band bending region of $55\nm$,
	providing an independent confirmation of the estimated 
	\Phigr.
At higher $B$, the diameter of the first QD state wave 
	function is dominated by $l_\mathrm{B}$ 
	rather than by the width of \Phigr 
	(Fig.~\ref{fig:Three_dot_states_V0_dependence}f).
The compression of the wave function for increasing $B$
	(Fig.~\ref{fig:Four_mag_field_dependency}f) 
	also manifests itself as increase in addition energy, 
	for instance, for
	$E_\mathrm{add}^1 = E_\mathrm{C}^1 + \Delta^\sigma$
	in Figure \ref{fig:Four_mag_field_dependency}d,
	where the increase in $E_\mathrm{add}^1$ with $B$ by 
	about $4\meV$ cannot be explained by that of 
	$\Delta^\sigma$, being $460\,\mathrm{\upmu eV}$ 
	between 3 and $7\T$.
Consequently, increased Coulomb repulsion between 
	electrons due to stronger compression and thus larger 
	$E_\mathrm{C}^1$ dominates $E_\mathrm{add}^1(B)$. 
We observe a similar monotonic increase 
	for the other $E_\mathrm{add}^i$ with odd index $i$,
	independent of the position of the QD.
\par
Experiment and theory also provide detailed insight
	into the valley splitting $\Delta^\tau_k$ of the 
	first confined states.
The peaks of the first quadruplets in 
	Figure~\ref{fig:One_topo_model_QD_evidence}a and 
	Figure~\ref{fig:Four_mag_field_dependency}a (see, 
	e.g., inset) often group in doublets, suggesting 
	sizable values of either $\Delta^\mathrm{\tau}_k$ or 
	$\Delta^\sigma$ (Fig.~\ref{fig:Zero}e,f).
While $\Delta^\sigma$ is expected to be spatially 
	homogeneous and only weakly varying between 
	different quadruplets, the TB calculations predict
	strongly varying $\Delta^\mathrm{\tau}_k$ for different
	quadruplets 
	(Fig.~\ref{fig:Three_dot_states_V0_dependence}e),
	in accordance with our observations in the
	experimental spectra. 
For a quantitative comparison we focus on 
	$E_\mathrm{add}^2$, which separates the two 
	doublets within the first quadruplet.
In view of the small value of the Zeeman splitting
	($\Delta^\sigma \approx 800\,\mathrm{\upmu eV}$
	at $7\T$), we  approximate $E^2_\mathrm{C}$ by $E^3_\mathrm{add}$ 
	to extract the valley splitting $\Delta^\tau_1
	\approx E^2_\mathrm{add}-E^3_\mathrm{add}$.
We record 20 spectra in the vicinity of an AA 
	stacked area at $B=7\T$ to obtain a histogram of 
	$\Delta^\tau_1$ for electrons and holes
	(Fig.~\ref{fig:Four_mag_field_dependency}e),
	where $\Delta^\tau_1$ could be determined with an 
	experimental error smaller than $0.2\meV$.
The values strikingly group around the predicted 
	$\Delta^\tau_1 \approx 3\meV$ found in the TB
	calculations
	(Fig.~\ref{fig:Three_dot_states_V0_dependence}e),
	with a probable offset in the QD position relative
	to the tunneling tip (\supp, Sec.~5) explaining
	the QD probing an area adjacent to the tunneling tip.
We conclude that sizable $\Delta^\tau_k$ separate 
	quadruplets into doublets, while the smaller
	$\Delta^\sigma$ contributes to the odd addition 
	energies within the	doublets.
Realizing such a controlled lifting of one of the two 
	degeneracies in graphene QDs is a key requirement 
	for 2-qubit gate operation \cite{Trauzettel2007}.
It enables Pauli blockade in exchange driven
	qubits as required for scalable quantum computation 
	approaches using graphene \cite{Trauzettel2007}.
Our observation of valley splittings, so far
	elusive, provides a stepping stone towards 
	the exploitation of the presumably large coherence 
	time of electron spins in graphene QDs
	\cite{Trauzettel2007,Fuchs2012,Droth2013,Fuchs2013}.
\par
%
In summary, we have realized graphene quantum dots
	without physical edges via electrostatic confinement 
	in magnetic field using low disorder graphene 
	crystallographically 
	aligned to a hexagonal boron nitride substrate.  
We observe more than 40 charging peaks in the hole 
	and electron sector arranged in quadruplets due
	to orbital splittings.
The first few peaks on the hole and electron side
	show an additional doublet structure traced 
	back to lifting of the valley degeneracy.
Note that such a lifting is key for the
	use of graphene quantum dots as spin qubits
	\cite{Trauzettel2007}.
Tight binding calculations quantitatively reproduce 
	the orbital splitting energy of $4-10\meV$
	as well as the first orbital's valley splitting 
	energy of about $3\meV$ by assuming a 
	tip potential deduced from an electrostatic Poisson 
	calculation.
Also the onset of confinement at about $2\T$
	is well reproduced by the calculation.
Our results demonstrate a much better controlled
	confinement 
	by combining magnetic and electrostatic 
	fields than previously found in graphene. 
Exploiting the present approach in transport merely
	requires replacing the tip by a conventional
	electrostatic gate with a diameter of about $100\nm$.
Moreover, the approach allows for straightforward tuning
	of (i) orbital splittings by changing the gate geometry
	and thus the confinement potential, 
	(ii) valley splittings based on substrate interaction, 
	(iii) the Zeeman splitting by altering the magnetic field, 
	and (iv) the coupling of dot states to leads
	or to other quantum dots by changing the magnetic field or
	selecting a different quantum dot state.
Finally, our novel mobile quantum dot enables a detailed
	investigation of structural details 
	of graphene stacked on various substrates, by spatially 
	mapping the quantum dot energies.
\par
%
The authors thank C.~Stampfer, R.~Bindel,
M.~Liebmann and K.~Fl\"ohr 
for prolific discussions, as well as C.~Holl for contributions 
to the Poisson calculations and A.~Georgi for 
assisting the measurements.
NMF, PN and MM gratefully acknowledge support from 
the Graphene Flagship (Contract No.~NECTICT-604391) and the 
German Science foundation (Li 1050-2/2 through SPP-1459).
LAC, JB and FL from the Austrian Fonds zur F\"orderung der 
wissenschaftlichen Forschung (FWF) through the SFB 041-ViCom 
and doctoral college Solids4Fun (W1243).
Calculations were performed on the Vienna Scientific Cluster.
RVG, AKG and KSN also acknowledge support from EPSRC 
(Towards Engineering Grand Challenges and Fellowship programs), 
the Royal Society, US Army Research Office, US Navy Research 
Office, US Airforce Research Office. KSN is also grateful to 
ERC for support via Synergy grant Hetero2D. AKG was supported 
by Lloyd’s Register Foundation.
\newpage
\bibliography{ESQD_arxiv}
\bibliographystyle{achemso}

\end{document}